\documentclass[twocolumn,preprintnumbers,showpacs,aps,
               amsmath,amsfonts]{revtex4}

\usepackage{graphicx}
\usepackage{epic,eepic}
\usepackage{amsthm}
\usepackage{hyperref}

\newtheorem{theorem}{Theorem}

\newtheorem{lemma}[theorem]{Lemma}

\newcommand{\lp}{\left}
\newcommand{\rp}{\right}
\newcommand{\be}{\begin{eqnarray}}
\newcommand{\ee}{\end{eqnarray}}
\newcommand{\beq}{\begin{equation}}
\newcommand{\eeq}{\end{equation}}
\newcommand{\ba}{\begin{array}}
\newcommand{\ea}{\end{array}}

\newcommand{\ket}[1]{{|{#1}\rangle}}
\newcommand{\bra}[1]{{\langle{#1}|}}

\newcommand{\mypmatrix}[1]{\begin{pmatrix}#1\end{pmatrix}}
\DeclareMathOperator{\diag}{diag}
\DeclareMathOperator{\Tr}{Tr}
\DeclareMathOperator{\RMS}{RMS}

\begin{document}


\preprint{CALT-68-2486}

\title{Quantum weak coin-flipping with bias of 0.192}

\author{Carlos Mochon}
\email{carlosm@theory.caltech.edu}
\affiliation{Institute for Quantum Information, 
California Institute of Technology,
Pasadena, CA 91125, USA}

\date{October 11, 2004}

\begin{abstract}
A family of protocols for quantum weak coin-flipping which asymptotically
achieve a bias of $0.192$ is described in this paper. The family contains 
protocols with $n+2$ messages for all $n>1$. The case $n=2$ is equivalent to 
the protocol of Spekkens and Rudolph with bias $1/\sqrt{2}-1/2 \simeq 0.207$. 
The case $n=3$ achieves a bias of $0.199$, and $n=8$ achieves a bias
of $0.193$. The analysis of the protocols uses Kitaev's
description of coin-flipping as a semidefinite program. The paper
constructs an analytical solution to the dual problem which provides an
upper bound on the amount that a party can cheat.
\end{abstract}



\maketitle

\section{Introduction}

Quantum coin-flipping is an attempt to solve the problem of coin-flipping
by telephone \cite{Blum} in a setting where the security is guaranteed purely
by the laws of quantum mechanics.

A typical description of the problem is as follows: Alice and Bob have been 
collaborating via  email and are about to publish. 
Given their notorious lack of surnames, they would like to flip a coin to 
determine whose name appears first on the paper. Unfortunately, though good 
collaborators, they don't completely trust  each other, nor do they 
have a common acquaintance whom they both trust. Though they don't wish to
meet in person, they do want to guarantee that the other person can't cheat
and win the coin-toss with a probability greater than $1/2$. Fortunately,
they have at their disposal computers capable of sending and processing
quantum email.

More specifically, quantum coin-flipping is a two party protocol involving a 
sequence of quantum messages between the parties, after which each party must 
output a classical bit. An output of zero will correspond to Alice winning
whereas Bob will win on an output of one. The requirements of the protocol 
are as follows:
(1) If both parties are honest then Alice's bit must be uniformly random, and
it must always equal Bob's bit; (2) If Alice is honest, then independently
of what Bob does, she will output one (i.e., Bob wins) with a probability
no greater than $P_B^* = \frac{1}{2} + \epsilon_B$; (3) If Bob is honest,
then independently of Alice's actions, he will output zero with a probability
no greater than $P_A^* = \frac{1}{2} + \epsilon_A$.
We define the bias as $\epsilon=\max(\epsilon_A,\epsilon_B)$, which is the
figure of merit for a coin-flipping protocol. In the ideal case we want
$\epsilon=0$. 

Note that no restriction is placed on the case when both players are dishonest.
Nor are there any requirements that the outcomes agree when one party is 
cheating, which would be impossible to achieve.
Furthermore, the protocol must start in an unentangled state, for if they could
start in a known entangled state of their choosing, the problem would be 
trivial.

Strictly speaking the above problem is given the name weak coin-flipping
because a cheating party may opt to lose. For example, no restriction is
placed on Bob's ability to force Alice to output zero. The case when neither
party may bias the coin in either direction is called strong coin-flipping.

Ambainis \cite{Ambainis2002} and Spekkens and Rudolph \cite{Spekkens2001}
have constructed strong coin-flipping protocols with a bias of
$\epsilon=\frac{1}{4}$. It is also known that quantum strong coin-flipping
protocols cannot achieve a bias smaller than $\epsilon=1/\sqrt{2}-1/2\simeq
0.207$, as was proven by Kitaev \cite{Kitaev} (and summarized in
Ref.~\cite{Ambainis2003}).

There is less known about weak coin-flipping. The best known protocol prior
to the present paper is by Spekkens and Rudolph \cite{Spekkens2002} and
achieves a bias of $\epsilon=1/\sqrt{2}-1/2\simeq 0.207$ (previous
protocols include Ref.~\cite{ker-nayak}). The best known lower bound is by
Ambainis \cite{Ambainis2002} and states that the number of rounds must grow
at least as $\Omega(\log \log \frac{1}{\epsilon})$.  In particular, this
means that no protocol having a fixed number of rounds can achieve an
arbitrarily small bias. Ambainis \cite{Ambainis2002bis} also proves 
the optimality of Spekkens and Rudolph's protocol within a family of 
3-message protocols.

It is likely that the bias achieved by Spekkens and Rudolph is optimal
for any protocol involving three or less messages. However, as we shall show
in this paper, a better bias can be achieved using more messages. 

In particular, we shall describe in Section~\ref{sec:proto} a family of
protocols, indexed by an integer $n>1$, with $n+2$ messages. The case of 
$n=2$ with four messages will be equivalent to Spekkens and Rudolph's
original protocol. The protocols with more rounds will achieve an even
better bias.

The main result of this paper is the construction of quantum weak coin-flipping
protocols which achieve a bias less than $\epsilon=1/\sqrt{2}-1/2$.
This excludes the possibility that Kitaev's bound for strong coin-flipping
can be directly extended to weak coin-flipping, and establishes that it is
not possible for the minimum bias of weak coin-flipping to equal that
for strong coin-flipping in the context of quantum mechanics.
We do not exclude the possibility, though, that a three
message protocol, outside of the family analyzed by Spekkens and Rudolph,
can achieve the optimal bias claimed by the present paper.

The main technique used in this paper is Kitaev's description of
coin-flipping as a semidefinite program. This description provides
a dual problem whose solutions bound the amount that a party may cheat.
Though this material has been previously published, it will be reviewed
in Section~\ref{sec:sdp}.

The main contribution of the present paper is in Section~\ref{sec:spec},
where we shall construct solutions to the problem dual to the protocol
of Section~\ref{sec:proto}. These shall provide analytic upper bounds
on the bias of the protocol.

Our protocol for a given $n$ depends on $n$ parameters
subject to one constraint, and we shall express the upper bound as a function
of these parameters. Any choice of the parameters, consistent with the
constraint, will give a valid protocol together with an upper bound on its 
bias. To find good protocols with small bias, we shall use
a numerical minimization over the space of parameters. This will be
done in Section~\ref{sec:num}.

We stress, however, that given a set of values for the parameters, these
can be put into the analytic expression to obtain a valid upper bound on the
bias. The existence of weak coin-flipping protocols with the quoted biases
does not depend in any way on the accuracy or quality of the numerical 
minimization.

Finally, the most interesting question is what happens in the limit
$n\rightarrow\infty$. We shall show that at least for some choices
of the parameters, the bias does not converge to zero. In fact, it appears
that most reasonable choices converge to the same point in this limit,
which may indicate that this is the best bias that can be achieved
by any quantum weak coin-flipping protocol.

\section{\label{sec:proto}The protocol}

We shall describe a family of weak coin-flipping protocols, indexed by an 
integer $n\geq 2$, which will involve $n+2$ messages. The protocols will also
depend on a set of parameters $a_1,\dots,a_n$ to be fixed later.
These parameters define the two-qubit states
\be
\ket{\phi_i} = \sqrt{a_i} \ket{00} + \sqrt{1-a_i} \ket{11}.
\ee

\noindent
The protocol begins with Alice preparing in her private Hilbert 
space the states $\ket{\phi_i}$ for odd $i$, while Bob prepares 
the states with even $i$ in his Hilbert space.

The first $n$ messages of the protocol consist of sending halves of the states
$\ket{\phi_i}$. More explicitly, the $i^{th}$ message involves the owner 
of state $\ket{\phi_i}$, who sends one of the two qubits comprising the state
to the other party. After the first $n$ messages, if both players were 
honest, the state of the system should be:
\beq
\ket{\psi}_{AB} = 
\bigotimes_{i=1}^n \lp( \sqrt{a_i} \ket{0}_A\otimes\ket{0}_B + 
\sqrt{1-a_i} \ket{1}_A\otimes\ket{1}_B \rp),
\eeq

\noindent
where the labels $A,B$ denote the owner of the qubit in question.

At this point each side will apply a two-outcome projective measurement 
$\{E_0,E_1\}$ to their $n$ qubits. These operators will be described below
but will have the properties $E_i^2 = E_i$ and 
$(E_i)_A \otimes I_B \ket{\psi}_{AB}=I_A\otimes (E_i)_B \ket{\psi}_{AB}$.
These properties guarantee that when both parties are honest, their 
answers are perfectly correlated. We can therefore associate the outcome 
$E_0$ with an outcome of zero for the coin flip, and the outcome $E_1$ 
with coin outcome one. The requirement that the coin-flip
be fair when both parties are honest:
\be
\bra{\psi} E_i \otimes E_i \ket{\psi} = \frac{1}{2}
\ee

\noindent
will impose a constraint on the parameters $\{a_i\}$.

At this point both parties should know the ``honest'' outcome of the coin-flip.
Now they enter a stage of cheat detection in which the loser will examine the
qubits of the winner. If no cheating is detected (which is guaranteed when
both players are honest) then the ``honest'' outcome becomes the final outcome.
Otherwise, if the losing party detects cheating, that party may ignore
the ``honest'' outcome and instead output his or her desired outcome
(zero for Alice and one for Bob).
This is acceptable because the rules of weak coin-flipping don't require
the parties to output the same bit when one party is dishonest.

We now describe the cheat detection stage which will involve the last two
messages: the winner of the coin toss according to the
measurement $\{E_0,E_1\}$ sends over their entire Hilbert space for inspection.
If Bob wins, he should send over his $n$ qubits so that Alice obtains both
halves of the state:
\be
\sqrt{2} E_1\otimes E_1 \ket{\psi} = \sqrt{2} E_1\otimes I \ket{\psi}.
\ee

\noindent
This is a pure state, and Alice can perform a two outcome projection
onto this state and its complement. If she obtains the complement as outcome,
she knows Bob must have cheated. More specifically, define
\be
F_i = \frac{E_i\otimes E_i \ket{\psi} \bra{\psi}E_i\otimes E_i}
{\bra{\psi} E_i\otimes E_i \ket{\psi}}.
\ee

\noindent
Alice measures using the projections $\{F_1,I-F_1\}$, where outcome $I-F_1$
implies Bob has cheated. In the case when the honest outcome is zero, Bob 
does the equivalent steps with $\{F_0,I-F_0\}$.

Officially, we shall define the
protocol so that Alice always uses message $n+1$ to either send her
qubits (or nothing if she lost the honest coin toss), whereas Bob will use
message $n+2$ if he needs to send qubits. The ordering is 
irrelevant though, and it could also be defined so that Bob sends his
verification qubits first when $n$ is odd, thereby avoiding two
messages in a row from Alice. Alternatively, they could be sent in the opposite
order, combining the verification state with the last message in order
to run the protocol with only $n+1$ messages.

All that remains is to describe the projections $E_0$ and $E_1$. Heuristically,
the measurement consists of the following process: Examine the qubits in
order starting from the one belonging to $\ket{\phi_n}$ and ending with
the one belonging to $\ket{\phi_1}$. The qubits are to be measured in
the computational basis, until the first zero outcome is obtained, which
implies that the sender of that qubit loses. If all qubits produce outcome
one then Alice (being the first message sender) is the winner. 
Of course, the measurement is not performed in
stages as described above but rather using the unique pair of projectors
which produces the same distribution of probabilities. For example, for
$n=2$ we have
\be
E_0 &=& \ket{00}\bra{00} + \ket{10}\bra{10} + \ket{11}\bra{11},
\\
E_1 &=& \ket{01}\bra{01},
\ee

\noindent
where the leftmost qubit corresponds to the first qubit sent or received.
The rest can be defined inductively by the formulas
\be
E_0^{(k+1)} &=& I \otimes E_1^{(k)} + \ket{1\cdots 1}\bra{1\cdots 1},
\\
E_1^{(k+1)} &=& I \otimes E_0^{(k)} - \ket{1\cdots 1}\bra{1\cdots 1},
\ee
\noindent
where the superscript indicates the number of qubits on which they are to
act. For brevity, these superscripts shall be omitted, though.

The case of $n=2$ is equivalent to the protocol described by
Spekkens and Rudolph in Ref.~\cite{Spekkens2002} which achieves the tradeoff 
$P_A^*P_B^*=1/2$. The connection is made by 
setting $a_1 = x$ and $(1-a_2) = 1/(2x)$.

Summarizing, the protocol involves the following steps:
\begin{enumerate}
\item 
Alice prepares 
$\ket{\phi_1}\otimes\ket{\phi_3}\otimes\ket{\phi_5}\cdots$,\\
Bob prepares $\ket{\phi_2}\otimes\ket{\phi_4}\otimes\ket{\phi_6}\cdots$.
\item For $i=1$ to $n$:\\
If $i$ is odd: Alice sends half of the state $\ket{\phi_i}$ to Bob,\\
If $i$ is even: Bob sends half of the state $\ket{\phi_i}$ to Alice.
\item 
Alice performs the two-outcome measurement $\{E_0,E_1\}$ on her $n$ qubits.
Bob performs the same two-outcome measurement $\{E_0,E_1\}$ on his $n$ qubits.
\item 
If Alice obtains $E_0$ she outputs zero, and sends all her qubits to Bob.
\item
If Bob obtains $E_1$ he outputs one, and sends all his qubits to Alice.
\item
If Alice obtained $E_1$ she measures her qubits plus any qubits received from
Bob with the projections $\{F_1,I-F_1\}$. If she obtains $F_1$ she outputs one,
otherwise (or if she receives the wrong number of qubits from Bob) she outputs
zero.
\item
If Bob obtained $E_0$ he measures his qubits plus any qubits received from
Alice with the projections $\{F_0,I-F_0\}$. If he obtains $F_0$ he outputs 
zero, otherwise (or if he receives the wrong number of qubits from Alice) he 
outputs one.
\end{enumerate}

\subsection{Reformulation of the protocol}

For the analysis in the following section, it will be helpful to delay
all measurements to the last step. It will be also useful
to never have to apply a unitary, or equivalently, send qubits conditioned
on the outcome of a measurement. We will be able to formulate protocols
with these properties if we are willing to allow one side to have
increased cheating power.

The idea is that the analysis of a coin-flipping protocol is divided into
two separate steps: we need to analyze the case when Alice is honest
and Bob is cheating, and then we need to analyze the case when Bob is
honest and Alice is cheating. Let us focus on the first case when Alice is
honest.

We wish to describe a new coin-flipping protocol, where Bob's
ability to cheat is exactly the same as in the original protocol, but where
Alice may be able to cheat more than usual. We shall call the original 
protocol $\mathcal{P}$ and the new protocol $\mathcal{P}'$. The idea is that
$\mathcal{P}'$ will be simpler to describe than $\mathcal{P}$ and since at
the moment we are only concerned with bounding Bob's ability to cheat,
any bound derived for one protocol will apply to the other.

The protocol $\mathcal{P}'$ begins with the same initial state as 
$\mathcal{P}$ and the first $n$ messages are identical. However, in 
$\mathcal{P}'$ after the first $n$ messages no measurements occur.
Instead, Bob sends all of his $n$ qubits to Alice. After this last message
Alice performs the two-outcome projective measurement $\{F_1,I-F_1\}$
as before and reports outcome $F_1$ as Bob winning and $I-F_1$ as Alice
winning. Note that in $\mathcal{P'}$, even when Bob is honest the outcome
$I-F_1$ can arise, that is, Alice does not differentiate between Bob losing
honestly and Bob getting caught cheating.

Technically, we should allow one last classical message from Alice to Bob,
where Alice announces her outcome and then Bob repeats it as his own, but
this won't be necessary as we are only concerned with the probabilities
associated with Alice's output.

It is not hard to see that any cheating strategy for Bob that can be
used in $\mathcal{P}$ will produce the same probability of winning
in $\mathcal{P}'$, because the only thing that changed from Alice's 
perspective is that now she always expects to receive Bob's qubits. However,
as protocol $\mathcal{P}$ was written, the only time that Bob could win 
was when he sent his qubits, so he loses nothing by always sending them.
From a mathematical perspective, we are using the fact that
$F_1 E_1 = F_1$ and $(I-F_1)E_1 + E_0 = I-F_1$.

In conclusion, when analyzing the case of honest Alice, we can use protocol 
$\mathcal{P}'$. Of course, when analyzing the case of honest Bob and cheating
Alice, $\mathcal{P}'$ is no longer useful, but we can define a new protocol
$\mathcal{P}''$ in a similar way, where Alice always sends all her qubits
to Bob. This protocol can be used to bound Alice's cheating power in 
$\mathcal{P}$.

For the rest of this paper, we shall employ protocols $\mathcal{P}'$ and
$\mathcal{P}''$ where appropriate without further comment. However,
all bounds derived will apply to the original protocol $\mathcal{P}$
as well.

\section{\label{sec:sdp}Coin-flipping as an SDP}

The problem of finding
the optimal cheating strategy for a player can be cast as a semidefinite 
program (SDP). The dual problem then provides bounds on the maximum bias
that the cheating player may achieve. This approach was first described by
Kitaev \cite{Kitaev} (and summarized in Ref.~\cite{Ambainis2003}).

In the following section we will review Kitaev's construction, though using
a somewhat different language than the original. What few results are needed
from the theory of semidefinite programming will be derived along the way
in order to keep this paper as self contained as possible. The discussion in
this section will be completely general in the sense that it applies to
any coin-flipping protocol. The results of this section will then be
applied to the protocol at hand in Section~\ref{sec:spec}.

\subsection{The primary problem}

For simplicity, we shall focus on the case when Alice is honest and Bob
is cheating. The opposite case when Bob is honest is nearly identical.

We will work with protocols that can be cast in the following form:
The initial state is a fixed pure unentangled state shared by Alice and
Bob. The protocol proceeds by applying unitaries on each individual side, and
by sending qubits from Alice to Bob and vice-versa. In the last step, each
party performs a two outcome projective measurement and outputs the result.

In fact, the communication part of the protocol (i.e.,
everything but the initial state preparation and the final measurement)
can be described as a sequence of the following three elementary operations:
one of Alice's qubits is sent to Bob, one of Bob's qubits is sent to Alice,
or each side applies a unitary to their qubits. The unitary
step is often not needed and can be completely removed if we allow
each party to decompose their space into qubits in different ways on each 
round.

Given a protocol, let $m$ be the number of elementary steps, and
let $\rho_0$ be the density matrix describing Alice's qubits in the first 
step. Let 
$\rho_i$ be the density matrix describing Alice's qubits after the first
$i$ elementary operations, given some cheating strategy for Bob. These
matrices must satisfy the following equations:
\begin{itemize}
\item If step $i$ involves sending qubit $j$ from Alice to Bob:
\be
\rho_{i} =\Tr_j \rho_{i-1}.
\ee
\item If step $i$ involves Alice receiving a qubit from Bob and assigning it
name $j$:
\be
\Tr_j \rho_{i} = \rho_{i-1}.
\ee
\item If step $i$ involves Alice applying unitary $U_i$:
\be
\rho_{i} = U_i \rho_{i-1} U_i^{-1}.
\ee
\end{itemize}
\noindent
It will be convenient to have a shorthand notation for these equations.
They shall be written as $L_i(\rho_i)=R_i(\rho_{i-1})$, where $L_i$ and $R_i$
are linear operators corresponding to the identity, partial trace, or
conjugation by a unitary as needed to match the above equations.

Clearly, no matter what Bob's strategy is, the above equations must be
satisfied. Furthermore, because Alice's output probabilities are entirely
determined by $\rho_m$, a cheating strategy for Bob can be described in
terms of the above sequence of density operators $\{\rho_i\}$. In fact,
it is not hard to see that by keeping the total state pure, Bob can make 
Alice have any sequence of density operators which are consistent with the 
above equations. Therefore, there is a one-to-one correspondence between
cheating strategies of Bob (up to isomorphisms that produce the same result 
on Alice's side) and density operators $\rho_0,\dots,\rho_m$ satisfying the
above equations. 

After the communication rounds have been completed, Alice makes a two-outcome 
projective measurement $\{E_A,E_B\}$ to determine her output. 
Outcome $E_A$ will correspond
to Alice winning (i.e., final outcome zero) and $E_B$ will correspond to to
Bob winning (i.e., final outcome one). Note that these operators are not
the same as the $\{E_0,E_1\}$ used in the last section, and in fact,
when applied to our protocol $E_B$ will correspond to $F_1$.

Bob's goal is to choose a sequence of positive semidefinite operators
$\rho_1,\dots,\rho_m$ satisfying the above protocol dependent equations, 
in order to maximize $\Tr(E_B \rho_m)$. Note that $\rho_0$ is always fixed by 
Alice's initial state, and the above equations fix the trace of the 
remaining matrices, therefore the maximization can indeed be done
over all positive semidefinite matrices. We have therefore proven the
following lemma:

\begin{lemma}
The maximum probability of winning that can be attained by Bob through cheating
in a coin-flipping protocol described by the data 
$m$, $\rho_0$, $\{L_i\}$, $\{R_i\}$, $E_B$ is given
by the solution of the maximization problem
\be
P_B^* = \max  \Tr \lp(E_B \rho_m \rp),
\label{eq:primin}
\ee
\noindent
involving the $m$ positive semidefinite matrices $\rho_1,\dots,\rho_m$ 
subject to the constraints
\be
L_i(\rho_i) = R_i(\rho_{i-1}) \quad \text{for all} \quad i=1,\dots,m.
\ee
\end{lemma}

\subsection{The dual problem}

The beauty of semidefinite programing is that each SDP has a dual SDP. When
the original problem involves a maximization, the dual problem involves
a minimization. Furthermore, the optimal solution of the dual problem will
be greater than or equal to the optimal maximum of the original problem.
In terms of coin-flipping each solution of the dual problem provides an 
upper bound on the amount that Bob can cheat.

The variables of a dual SDP are Lagrange multipliers, one for each constraint
in the original problem. There are $m$ equality constraints given by the
$m$ elementary operations of the protocol, therefore there will be
$m$ Lagrange multipliers $Z_1,\dots,Z_m$. Each $Z_i$ will be a Hermitian
matrix of the same dimension as $L_i(\rho_i)$ and will be added in as
a term of the form $\Tr[Z_i (L_i(\rho_i) - R_i(\rho_{i-1}))]$.

We will now lift the conditions $L_i(\rho_i) = R_i(\rho_{i-1})$ on the 
operators $\{\rho_i\}$ allowing them to vary freely. The constraints
will be dynamically imposed by the Lagrange multiplier terms. However,
because the traces of $\{\rho_i\}$ are no longer fixed we shall impose the
constraints $\rho_i\leq I$ so as to keep the expression $\Tr E_B \rho_m$
finite. We now have:
\begin{widetext}
\be
P_B^* &=& \underset{0\leq\rho_1,\dots,\rho_m\leq I}{\max} \lp\{  \Tr E_B \rho_m
- \underset{Z_1,\dots,Z_m}{\sup} \sum_{i=1}^m 
\Tr\lp[Z_i (L_i(\rho_i) - R_i(\rho_{i-1})) \rp] \rp\}
\nonumber\\
&=& \underset{0\leq\rho_1,\dots,\rho_m\leq I}{\max} \ 
\underset{Z_1,\dots,Z_m}{\inf} \lp\{
\Tr E_B \rho_m - \sum_{i=1}^m \Tr Z_i L_i(\rho_i) 
+ \sum_{i=1}^m \Tr Z_i R_i(\rho_{i-1}) \rp\}
\nonumber\\
&=& \underset{0\leq\rho_1,\dots,\rho_m\leq I}{\max} \ 
\underset{Z_1,\dots,Z_m}{\inf} \lp\{
\Tr Z_1 R_1(\rho_0) + \sum_{i=1}^m 
\lp( \Tr R_{i+1}(\rho_i) Z_{i+1} - \Tr L_i(\rho_i) Z_i  \rp) \rp\}
\nonumber\\
&=& \underset{0\leq\rho_1,\dots,\rho_m\leq I}{\max} \ 
\underset{Z_1,\dots,Z_m}{\inf} \lp\{
\Tr Z_1 R_1(\rho_0) + \sum_{i=1}^m 
\Tr \lp[ \rho_i \lp(R_{i+1}^d( Z_{i+1}) - L_i^d(Z_i) \rp)\rp]\rp\},
\ee
\end{widetext}

\noindent
where in the third line we introduced $Z_{m+1} \equiv E_B$ and 
$R_{m+1}(\rho)=\rho$. In the fourth line, we introduced the dual operators
to $L_i$ and $R_i$ in the sense that $\Tr[L_i(\rho) Z]= \Tr[\rho L_i^d(Z)]$
and $\Tr[R_{i}(\rho) Z]= \Tr[\rho R_i^d(Z)]$ for all $\rho$ and $Z$.
These are easily constructed as follows: if $R_i(\rho) = \rho$ then 
$R_i^d(Z) = Z$, if $R_i(\rho) = U_i \rho U_i^{-1}$ then 
$R_i^d(Z) = U_i^{-1} Z U_i$, and if 
$R_i(\rho) = \Tr_j \rho$ then $R_i^d(Z) = Z \otimes I_j$, where the
identity is inserted into the empty slot of qubit $j$. The expressions
for $L_i^d$ are defined similarly.

From the above equation it should be clear that
\be
P_B^* \leq \Tr[Z_1 R_1(\rho_0)],
\ee
\noindent
for any Hermitian matrices $Z_1,\dots,Z_m$ subject to the $m$ constraints
\be
R_{i+1}^d( Z_{i+1}) - L_i^d(Z_i) \leq 0,
\ee
\noindent
because under this constraint the second term is guaranteed to be non-positive
for any set of $\{\rho_i\}$. We have therefore proven the following
theorem:

\begin{theorem}
Let $Z_1,\dots,Z_m$ be any set of Hermitian matrices satisfying the $m$
inequalities
\be
L_i^d(Z_i) \geq R_{i+1}^d( Z_{i+1}) \quad \text{for} \quad i=1,\dots,m,
\label{eq:dualineq}
\ee

\noindent
where $m$, $L_i^d$, $R_i^d$, $Z_{m+1}\equiv E_B$, and $\rho_0$ are data 
associated with a coin-flipping protocol. The maximum probability 
that Bob can win such a coin-flip by cheating is bounded by
\be
P_B^* \leq \Tr[Z_1 R_1(\rho_0)].
\ee
\end{theorem}

Our goal in the next section will be to guess sets of matrices
$Z_1,\dots,Z_m$ satisfying the inequalities Eq.~(\ref{eq:dualineq}),
and try to find a set that produces a good bound on $P_B^*$ without
worrying whether the bound is optimal.

\section{\label{sec:spec}Finding solutions to the dual problem}

Continuing the analysis of the case where Alice is honest and Bob is cheating,
we need to find the problem dual to $\mathcal{P}'$. The protocol $\mathcal{P}'$
can be thought of as having $m=n+1$ elementary operations if we relax the 
definition somewhat to allow the receiving of $n$ qubits in the last message
as one step. Each elementary step consists of either sending or receiving
a message, and unitaries are never used. The final measurement is done with
$E_B=F_1$, $E_A=I-F_1$.

It will be useful to define a specific ordering for the qubits in Alice's
Hilbert space. The intuition is to picture qubits as carried by particles
in a lattice. When played honestly, the initial state will be prepared on
$2n$ particles, some of which will be controlled by Alice, and some by Bob.
Sending a qubit from Alice to Bob simply means that the particle will now
be controlled by Bob rather than Alice. Alice's full Hilbert space
at each step will be the ordered tensor product of the Hilbert spaces
of all particles she controls in that step. Note that this does not
restrict the power of a cheating player, who could have as many extra qubits
as he wants that can interact with any particle under his control.

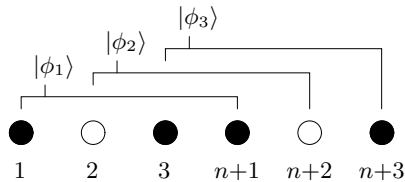
\begin{figure}
\setlength{\unitlength}{0.00083333in}
\begin{picture}(2616,1175)(-100,-35)
\put(1433,300){\blacken\ellipse{150}{150}}
\put(1433,300){\ellipse{150}{150}}
\put(83,300){\blacken\ellipse{150}{150}}
\put(83,300){\ellipse{150}{150}}
\put(533,300){\ellipse{150}{150}}
\put(983,300){\blacken\ellipse{150}{150}}
\put(983,300){\ellipse{150}{150}}
\put(1883,300){\ellipse{150}{150}}
\put(2333,300){\blacken\ellipse{150}{150}}
\put(2333,300){\ellipse{150}{150}}
\path(83,450)(83,525)(1433,525)(1433,450)
\path(233,525)(233,600)
\path(533,600)(533,675)(1883,675)(1883,450)
\path(683,675)(683,750)
\path(983,750)(983,825)(2333,825)(2333,450)
\path(1133,825)(1133,900)
\put(158,675){$\ket{\phi_1}$}
\put(608,825){$\ket{\phi_2}$}
\put(1058,975){$\ket{\phi_3}$}
\put(38,20){1}
\put(488,20){2}
\put(938,20){3}
\put(1288,20){$n$+1}
\put(1738,20){$n$+2}
\put(2188,20){$n$+3}
\end{picture}
\caption{State ordering for $n=3$. Black qubits are initially prepared by
Alice.}
\label{fig:particles}
\end{figure}

The ordering of the states will be as follows. The initial state is prepared
so that $\ket{\phi_i}$ is carried by particles $i$ and $n+i$. When $n$ is even
Alice starts off with all the odd particles in her possession
whereas Bob has all the even particles. When $n$ is odd Alice owns the
odd particles between $1$ and $n$ inclusive, and the even particles between
$n+1$ and $2n$ inclusive. This is depicted in Fig.~\ref{fig:particles}.

The first steps involve Alice sending
qubit $n+1$, then receiving qubit $2$, then sending qubit $n+3$, and so on.
At the end of the first $n$ messages Alice will control the first $n$
qubits. The last step involves Alice taking possession of the other $n$ qubits.

With these conventions, the primal problem reads:
\be
\rho_i &=& \Tr_{n+i} \rho_{i-1} \quad \text{for odd}\  i\leq n,\\
\Tr_{i} \rho_i &=& \rho_{i-1} \quad\quad\quad\ \ \text{for even}\  i\leq n,
\ee
\noindent
plus one final equation
\be
\Tr_{n+1,\dots,2n} \rho_{n+1} = \rho_n.
\ee

With these conventions the dual problem involves finding $m=n+1$ Hermitian 
matrices $Z_1,\dots,Z_{n+1}$. When $n$ is odd, all the matrices
have dimension $2^n$, whereas when $n$ is even the matrix $Z_{n+1}$ has
dimension $2^n$ and the rest have dimension $2^{n-1}$. They must satisfy
the following equations
\be
Z_i &\geq& Z_{i+1} \quad\quad\quad\quad\quad\, \text{for odd}\  i\leq n,\\
Z_{i}\otimes I_i &\geq& Z_{i+1} \otimes I_{n+i+1} \quad\text{for even}\ i<n,
\ee

\noindent
where the subscript on the qubit 
identity matrices indicate into which slot it should be inserted. 
If $n$ is even, we also need $Z_n \otimes I_n \geq Z_{n+1}$.
Finally, in addition to the previous $n$ inequalities we need to satisfy
\be
Z_{n+1} \otimes I_{n+1,\dots,2n} \geq Z_{n+2} \equiv F_1,
\ee

\noindent
where the identity is inserted into the slot of the last $n$ qubits. The goal
is to choose the matrices in order to minimize 
$\bra{\varphi} Z_1 \otimes I_{n+1} \ket{\varphi}$
where $\ket{\varphi} = \ket{\phi_1}\otimes\ket{\phi_3}\otimes\cdots$.

\subsection{Choosing $Z_1,\dots,Z_n$}

Let $\beta = \bra{\varphi} Z_1 \otimes I_{n+1} \ket{\varphi}$. 
To minimize this quantity, it is to our advantage to choose the 
$Z_i$ matrices as small as possible in a sense to be discussed below.
In particular, the optimal choice for 
$Z_1$ is simply to satisfy the equality $Z_1 = Z_2$. We can 
remove $Z_1$ from our equations and write
\be
\beta = \bra{\varphi} Z_2 \otimes I_{n+1} \ket{\varphi} 
= \bra{\varphi_3} \Tr_{a_1} Z_2 \ket{\varphi_3},
\ee

\noindent
where $\ket{\varphi_3}=\ket{\phi_3}\otimes\ket{\phi_5}\otimes\cdots$, and
$\Tr_{a_1}$ denotes a weighted partial trace on the first qubit
with weights $a_1$ and $1-a_1$. For example, when acting on a matrix
that only involves the first qubit
\be
\Tr_{a_1} M = a_1 \bra{0}M\ket{0} + (1-a_1) \bra{1}M\ket{1}.
\ee
\noindent
Note that in a slight abuse of notation, the subscript 1 in $\Tr_{a_1}$
indicates both which $a_i$ is used, and on which qubit the partial trace
is performed.

The next inequality, which reads $Z_2 \otimes I_2 \geq Z_3 \otimes I_{n+3}$,
is harder to satisfy, and in general equality cannot
be achieved. However, we don't need to pay much attention to what
happens in the subspace orthogonal to $\ket{\varphi_3}$, and we can in
a sense sacrifice this subspace in order to obtain small entries in the 
subspace that we are interested in.

More specifically, let $T_3$ be the partial trace
\be
T_3(M)=\Tr_{\ket{\varphi_3}}[(I_{1,2}\otimes\ket{\varphi_3}\bra{\varphi_3})M],
\ee
\noindent
where the trace is taken only over qubits that are involved in 
$\ket{\varphi_3}$, that is, qubits 3, $n+3$, 5, $n+5$, and so on.
The equation $Z_2 \otimes I_2 \geq Z_{3} \otimes I_{n+3}$ requires 
$T_3(Z_2 \otimes I_2) \geq T_3(Z_3 \otimes I_{n+3})$, which is an equation 
involving only the first two qubits. We will begin by finding the optimal
choice in this subspace.

Let us assume that $T_3(Z_3 \otimes I_{n+3})$ is a diagonal matrix with 
entries $x_{00},x_{01},x_{10},x_{11}$. 
We want to choose $T_3(Z_2 \otimes I_2)$ to be 
as small as possible while still satisfying the inequality. However,
because of the linearity of $T_3$, the matrix $T_3(Z_2 \otimes I_2)$
will have the form $M\otimes I_2$, for some one-qubit operator $M$. 
The equation $M\otimes I_2 \geq T_3(Z_3 \otimes I_{n+3})$ becomes
\beq
\mypmatrix{M_0 && 0 && M_{c} && 0 \cr 0 && M_0 && 0 && M_{c} \cr
           M_{c}^* && 0 && M_1 && 0 \cr 0 && M_{c}^* && 0 && M_1 }
\geq \mypmatrix{x_{00} && 0 && 0 && 0 \cr 0 && x_{01} && 0 && 0 \cr
                0 && 0 && x_{10} && 0 \cr 0 && 0 && 0 && x_{11} },
\eeq

\noindent
where $M_0$, $M_1$ are the diagonal entries of $M$ in the computational basis,
and $M_{c}$ is the complex off-diagonal entry.

Since we are trying to minimize $\bra{\varphi_3} \Tr_{a_1} Z_2 \ket{\varphi_3}
= \Tr_{a_1} M$, the best choice is to take $M_0=\max(x_{00},x_{01})$,
$M_1=\max(x_{10},x_{11})$ and $M_{c}=0$ which clearly satisfies the
inequality. Notice that the maximum is taken over pairs of eigenvalues
whose computational basis eigenvectors differ only in the second qubit.
Symbolically, we shall write this as
\be
M = \underset{2}{\max}\lp[ T_3(Z_3 \otimes I_{n+3})\rp],
\ee

\noindent
where the operator $\max$ is defined only for diagonal matrices.
The subscript $2$ specifies that the maximum is to be taken over
subspaces that differ in the second qubit.

The above discussion is only valid when $T_3(Z_3 \otimes I_{n+3})$ is
diagonal, but we can impose this constraint on $Z_3$ (and the
equivalent constraint on future $Z_i$), which is acceptable because
we are only looking for a solution of the inequalities, even if it is
not the optimal solution.

Now if we could choose $Z_2$ to satisfy the full inequality, and still
satisfy $T_3(Z_2)=M$ for the matrix chosen above we would have
\be
\beta = \Tr_{a_1} \underset{2}{\max} \lp[ T_3(Z_3 \otimes I_{n+3})\rp].
\ee

The following lemma shows that it is possible to choose $Z_2$ so that we can
get arbitrarily close to the above result. Because we will use $\beta$ to
upper bound $P_B^*$, it doesn't matter if it is an infimum, and therefore
we can use the lemma to eliminate $Z_2$ in favor of the above expression.

\begin{lemma}
Let $T_3$ be as above, and let $H$ be a Hermitian matrix with finite
eigenvalues.
Given a Hermitian matrix $M$ such that $M\otimes I_2 \geq  T_3(H)$
and an $\epsilon>0$, there exists a matrix $M'$ such that 
$M'\otimes I_2 \geq H$ and $T_3(M') = M + \epsilon I$.
\end{lemma}

\begin{proof}
Let $P = I_{1,2}\otimes\ket{\varphi_3}\bra{\varphi_3}$ be the projector
which was used in defining $T_3$. This divides the Hilbert space
on which $H$ acts into the direct sum of two parts, one invariant under
$P$ and one perpendicular to it. We write this as 
$\mathcal{H} = \mathcal{H}^\parallel \oplus \mathcal{H}^\perp$.

Let $\lambda$ be the largest eigenvalue of $(I-P)H(I-P)$. Define the
block diagonal matrix $B$ as follows: the block acting on the space
$\mathcal{H}^\parallel$ has the form $M\otimes I_2 + \epsilon I$,
and the block acting on $\mathcal{H}^\perp$ has the form $(\lambda + y) I$
for some constant $y>0$.

Let $\gamma$ be the maximum over normalized states $\ket{\Psi}$,$\ket{\Phi}$ of
$|\bra{\Phi} P H (I-P) \ket{\Psi}|$. Then for any normalized state
$\ket{\Psi}$ we have
\be
\bra{\Psi} ( B - H ) \ket{\Psi} &\geq& 
\epsilon \bra{\Psi}P\ket{\Psi}+ y \bra{\Psi}(I-P)\ket{\Psi}
\\\nonumber
&&- 2 \gamma \sqrt{\bra{\Psi}P\ket{\Psi} \bra{\Psi}(I-P)\ket{\Psi}}.
\ee

\noindent
As long as $y> \frac{\sqrt{\gamma}}{\epsilon}$, the expression is
greater than zero, which implies $B>H$. It should be clear that
$B$ has the form $M'\otimes I_2$ and that the $M'$ defined in this
way satisfies $T_3(M') = M + \epsilon I$.
\end{proof}

The above lemma is used with $H=Z_3 \otimes I_{n+3}$, and letting $Z_2 = M'$.
At this point the pattern begins to repeat itself. We can choose
$Z_3 = Z_4$ and get
\beq
\beta = \Tr_{a_1} \underset{2}{\max} \lp[ T_3(Z_4 \otimes I_{n+3}) \rp]
= \Tr_{a_1} \underset{2}{\max} \lp[ \Tr_{a_3} T_5(Z_4) \rp],
\eeq

\noindent
where $T_5$ is the partial trace using the
state $\ket{\varphi_5}=\ket{\phi_5}\otimes\ket{\phi_7}\otimes\cdots$. 

Using the lemma again we eliminate $Z_4$ in favor of $Z_5$:
\be
\beta = \Tr_{a_1} \underset{2}{\max} \lp[ \Tr_{a_3} 
\underset{4}{\max}\lp[T_5(Z_5\otimes I_{n+5})\rp] \rp],
\ee

\noindent
where the expression is only valid if $T_5(Z_5\otimes I_{n+5})$ is diagonal 
in the computational basis (which will force $T_3(Z_3\otimes I_{n+3})$ to 
be diagonal as well).

One may worry that repeated uses of the lemma will make $Z_3$ have arbitrarily
large entries which means that the lemma can no longer be used to eliminate
$Z_2$. But the problems can be eliminated by taking the limits in the
proper order, or more appropriately, by making sure that the coefficient
$y$ associated with $Z_2$ is much larger than the one associated with $Z_4$
which in turn needs to be much larger than the one associated with
$Z_6$ and so on.

The process is repeated until in the last step, when $n$ is odd,
the innermost expression is of the form 
$T_n(Z_n\otimes I_{2n}) = T_n(Z_{n+1}\otimes I_{2n}) = \Tr_{a_n} Z_{n+1}$,
yielding
\be
\beta = \Tr_{a_1} \underset{2}{\max} \lp[ \Tr_{a_3} \underset{4}{\max}
\lp[ \Tr_{a_5}
\cdots \Tr_{a_n} \lp[ Z_{n+1} \rp] \rp] \rp].
\label{eq:summax1}
\ee
\noindent
When $n$ is even, we had the special inequality $Z_n \otimes I_n \geq Z_{n+1}$
which is satisfied by choosing $Z_n = \underset{n}{\max}[Z_{n+1}]$, so that
we get the same alternating expression, with the innermost operation a $\max$:
\be
\beta = \Tr_{a_1} \underset{2}{\max} \lp[ \Tr_{a_3} \underset{4}{\max}
\lp[ \Tr_{a_5}
\cdots \underset{n}{\max} \lp[ Z_{n+1} \rp] \rp] \rp].
\ee
\noindent
Both of these formulas are valid only
if $Z_{n+1}$ is diagonal in the computational basis, which will make
all the matrices of the form $T_i(Z_i)$ for odd $i$ diagonal as well.

We are now left with the task of minimizing $\beta$ as a function of $Z_{n+1}$
with the constraint that $Z_{n+1}$ must be real and diagonal in the 
computational basis and must satisfy the inequality  
$Z_{n+1} \otimes I \geq F_1$.

In fact, when $Z_{n+1}$ is diagonal the inequality can be simplified further.
In the qubit ordering we have chosen, the final state of the protocol right
before measurement should be 
$\ket{\psi} = \ket{\phi_1}_{1,n+1}\otimes\ket{\phi_2}_{2,n+2}
\otimes\cdots\otimes\ket{\phi_n}_{n,2n}$ where we have explicitly listed the 
location of each qubit. Therefore 
$F_1 =  2 E_1\otimes E_1 \ket{\psi}\bra{\psi}E_1\otimes E_1$
has support only on the $2^n$ dimensional subspace spanned by states where 
qubits $i$ and $i+n$ are equal for all $i$. The constraint 
$Z_{n+1} \otimes I \geq F_1$ need only be checked in
this subspace where it takes the form
\be
Z_{n+1}\geq \ket{\xi_B}\bra{\xi_B},
\ee

\noindent
where
\be
\ket{\xi} &=& 
\lp( \sqrt{a_1} \ket{0} + \sqrt{1- a_1} \ket{1} \rp) 
\otimes \lp( \sqrt{a_2} \ket{0} + \sqrt{1- a_2} \ket{1} \rp)
\nonumber\\
&& \cdots \otimes\lp( \sqrt{a_n} \ket{0} + \sqrt{1- a_n} \ket{1} \rp),
\ee

\noindent
and $\ket{\xi_B} = \sqrt{2} E_1 \ket{\xi}$, which is correctly normalized 
by the factor $\sqrt{2}$ if the coin is fair when both players are honest.

\subsection{Example $n=3$}

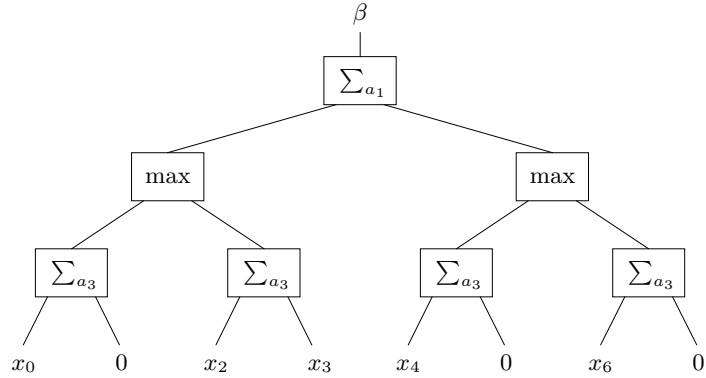
\begin{figure*}[!tb]
\setlength{\unitlength}{0.00083333in}
\begin{picture}(4292,2400)(0,-10)
\path(600,525)(150,525)(150,825)
	(600,825)(600,525)
\path(225,525)(75,225)
\path(525,525)(675,225)
\path(1800,525)(1350,525)(1350,825)
	(1800,825)(1800,525)
\path(1425,525)(1275,225)
\path(1725,525)(1875,225)
\path(3000,525)(2550,525)(2550,825)
	(3000,825)(3000,525)
\path(2625,525)(2475,225)
\path(2925,525)(3075,225)
\path(4200,525)(3750,525)(3750,825)
	(4200,825)(4200,525)
\path(3825,525)(3675,225)
\path(4125,525)(4275,225)
\path(1200,1125)(750,1125)(750,1425)
	(1200,1425)(1200,1125)
\path(3600,1125)(3150,1125)(3150,1425)
	(3600,1425)(3600,1125)
\path(375,825)(825,1125)
\path(1125,1125)(1575,825)
\path(2775,825)(3225,1125)
\path(3525,1125)(3975,825)
\path(2400,1725)(1950,1725)(1950,2025)
	(2400,2025)(2400,1725)
\path(975,1425)(2025,1725)
\path(2325,1725)(3375,1425)
\path(2175,2175)(2175,2025)
\put(2130,2250){$\beta$}
\put(2045,1838){$\sum_{a_1}$}
\put(835,1233){$\max$}
\put(3235,1233){$\max$}
\put(245,638){$\sum_{a_3}$}
\put(1445,638){$\sum_{a_3}$}
\put(2645,638){$\sum_{a_3}$}
\put(3845,638){$\sum_{a_3}$}
\put(0,70){$x_0$}
\put(650,70){$0$}
\put(1200,70){$x_2$}
\put(1850,70){$x_3$}
\put(2400,70){$x_4$}
\put(3050,70){$0$}
\put(3600,70){$x_6$}
\put(4250,70){$0$}
\end{picture}
\caption{Cheating Bob's Sum-Max Tree}
\label{fig:beta}
\end{figure*}

At this point an example would probably be helpful. We shall look at the
case $n=3$:
\be
\ket{\xi_B} = \sqrt{2} &\big(& 
\sqrt{a_1 a_2 a_3} \ket{000} 
+ \sqrt{a_1 (1-a_2) a_3} \ket{010} \nonumber\\&&
+ \sqrt{a_1 (1-a_2) (1-a_3)} \ket{011} \nonumber\\&&
+ \sqrt{(1-a_1) a_2 a_3} \ket{100}  \nonumber\\&&
+ \sqrt{(1-a_1) (1-a_2) a_3} \ket{110} \big),
\ee

\noindent
and the matrix $Z_{4}$ can be chosen as
\be
Z_{4} = \mypmatrix{ x_0 & 0   & 0   & 0   & 0   & 0   & 0   & 0    \cr
                    0   & 0   & 0   & 0   & 0   & 0   & 0   & 0    \cr
                    0   & 0   & x_2 & 0   & 0   & 0   & 0   & 0    \cr
                    0   & 0   & 0   & x_3 & 0   & 0   & 0   & 0    \cr
                    0   & 0   & 0   & 0   & x_4 & 0   & 0   & 0    \cr
                    0   & 0   & 0   & 0   & 0   & 0   & 0   & 0    \cr
                    0   & 0   & 0   & 0   & 0   & 0   & x_6 & 0    \cr
                    0   & 0   & 0   & 0   & 0   & 0   & 0   & 0    }
\ee

\noindent
where the top row corresponds to $\ket{000}$, the second one to $\ket{001}$
and so on. The entries along the diagonal of $Z_4$ outside of the support of 
$\ket{\xi_B}\bra{\xi_B}$ have already been set to zero, which should
be expected for all optimal solutions. Otherwise, any set of $\{x_i\}$ that
satisfies $Z_4\geq\ket{\xi_B}\bra{\xi_B}$ is a valid solution of the
dual problem. The corresponding bound can be calculated from these variables
using Eqs.~(\ref{eq:summax1}) or equivalently by evaluating 
the tree depicted in Fig.~\ref{fig:beta}.

The tree is evaluated as follows: each node has a value that is either
the maximum or the weighted sum of the nodes below it. The weighted sum
is just $a_i$ times the value of the left descendant plus $1-a_i$ times the
value of the right descendant. The value of the root node corresponds to 
$\beta$ and is an upper bound on $P_B^*$. We shall call trees of this form
Sum-Max trees.

Sum-Max trees appear naturally when analyzing classical protocols for
coin-flipping. The basic idea is that these protocols can be described as a 
sequence of public random bits, with the first one announced by Alice, then the
second one by Bob and so on. At the end both parties look at the sequence
of bits and determine the outcome of the coin-flip. The whole protocol
can be described as a binary tree, with Alice's bits choosing the path
at the odd depth nodes and Bob's bits controlling the rest.

A player attempting to cheat in such a protocol will not output random
bits but will instead choose the path that maximizes his chances of
winning at each node. If we put ones and zeros in the leaf nodes corresponding
to a win or loss, the maximum probability with which the cheater can
win is given by evaluating the corresponding Sum-Max tree.

If we ignore the cheat detection stage in the protocol, then it can be
described completely classically. Its Sum-Max tree would be the same as 
Fig.~\ref{fig:beta}, except that all the variables would be replaced by the 
number one. This can easily be seen from our formalism, because the only effect
of removing the last round is to force $Z_{n+1}$ to equal $E_1$ which is
diagonal with ones in place of the variables $\{x_i\}$.

It is well known that in the classical case, one party can always fully
bias the coin in their favor. However, in the quantum case with cheat 
detection, because the leaves of Sum-Max tree are less restricted,
there is the possibility of obtaining a stronger bound on the amount
of cheating.

The analysis so far has been of the case when Alice is honest and Bob is
cheating. The case of Bob honest and Alice cheating is almost identical, 
though.
The first major difference, is that this case has to be analyzed from
Bob's perspective, so that the odd messages consist of receiving a qubit
and the even ones involve sending a qubit. This has the effect of switching
sums with maxes and vice versa. The other difference is the final state
that Bob will use to verify that Alice is not cheating. For $n=3$ it has
the form
\be
\ket{\xi_A} = \sqrt{2} &\big(& 
\sqrt{a_1 a_2 (1-a_3)} \ket{001} \\\nonumber&&
+ \sqrt{(1-a_1) a_2 (1-a_3)} \ket{101}  \\\nonumber&&
+ \sqrt{(1-a_1) (1-a_2) (1-a_3)} \ket{111} \big).
\ee

\begin{figure*}[!tb]
\setlength{\unitlength}{0.00083333in}
\begin{picture}(4292,2400)(0,-10)
\path(600,525)(150,525)(150,825)
	(600,825)(600,525)
\path(225,525)(75,225)
\path(525,525)(675,225)
\path(1800,525)(1350,525)(1350,825)
	(1800,825)(1800,525)
\path(1425,525)(1275,225)
\path(1725,525)(1875,225)
\path(3000,525)(2550,525)(2550,825)
	(3000,825)(3000,525)
\path(2625,525)(2475,225)
\path(2925,525)(3075,225)
\path(4200,525)(3750,525)(3750,825)
	(4200,825)(4200,525)
\path(3825,525)(3675,225)
\path(4125,525)(4275,225)
\path(1200,1125)(750,1125)(750,1425)
	(1200,1425)(1200,1125)
\path(3600,1125)(3150,1125)(3150,1425)
	(3600,1425)(3600,1125)
\path(375,825)(825,1125)
\path(1125,1125)(1575,825)
\path(2775,825)(3225,1125)
\path(3525,1125)(3975,825)
\path(2400,1725)(1950,1725)(1950,2025)
	(2400,2025)(2400,1725)
\path(975,1425)(2025,1725)
\path(2325,1725)(3375,1425)
\path(2175,2175)(2175,2025)
\put(2130,2250){$\alpha$}
\put(2035,1833){$\max$}
\put(845,1238){$\sum_{a_2}$}
\put(3245,1238){$\sum_{a_2}$}
\put(235,633){$\max$}
\put(1435,633){$\max$}
\put(2635,633){$\max$}
\put(3835,633){$\max$}
\put(0,70){$0$}
\put(650,70){$x_1$}
\put(1200,70){$0$}
\put(1850,70){$0$}
\put(2400,70){$0$}
\put(3050,70){$x_5$}
\put(3600,70){$0$}
\put(4250,70){$x_7$}
\end{picture}
\caption{Cheating Alice's Sum-Max Tree}
\label{fig:alpha}
\end{figure*}
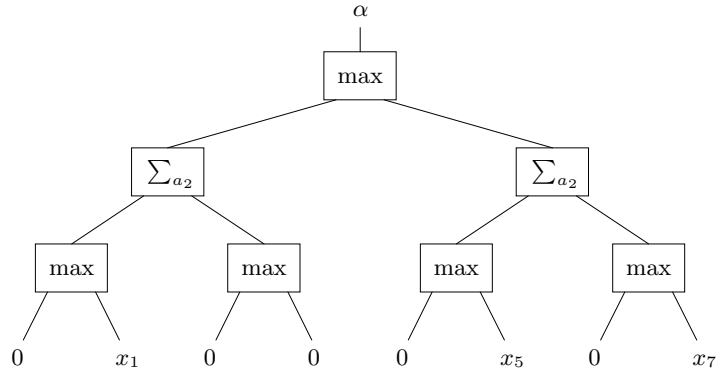

The maximum probability with which Alice can cheat, $P_A^*$, is bounded
above by $\alpha$, calculated from the Max-Sum tree in Fig.~\ref{fig:alpha},
where the leaves are the diagonal elements of $Z_4$ and must satisfy
$Z_4\geq \ket{\xi_A}\bra{\xi_A}$.

\subsection{Finding the optimal $Z_{n+1}$}

Returning to the case of honest Alice, we need to finish the general
case by choosing a matrix $Z_{n+1}$ in order to obtain an expression
for $\beta$ in terms of the parameters $a_1,\dots,a_n$.
Recall that we have restricted our analysis to matrices $Z_{n+1}$ that are
diagonal in the computational basis. We shall now search for the minimum
value of $\beta$ consistent with this choice.

Let $x_1,\dots,x_{2^n}$ be the diagonal entries of $Z_{n+1}$. We will,
as in the example above, set the variables to zero when their corresponding
basis vector is orthogonal to $\ket{\xi_B}$, which will leave
around half of the variables. We also wish to work in the subspace
where the two values entering a $\max$ node in the Sum-Max tree are equal.
For example, this is the space consistent with 
$a_3 x_0 = a_3 x_2 + (1-a_3) x_3$
and $x_4=x_6$ in the example above. The only potential problem exists
at the lowest level of $\max$ nodes, where a zero can be entering the node.
For the following we will assume that $n$ is odd which eliminates this problem.
The even case will be derived from the odd case below.

Working in this subspace we can replace all the maximums with weighted sums
with any weight of our choice. In this situation, $\beta$ can be calculated
as a weighted trace of $Z_{n+1}$. That is, there exist diagonal matrices $W$
such that $\beta=\Tr(Z_{n+1} W)$ for any $Z_{n+1}$ in this subspace.
For example, a valid choice for $W$ is
the diagonal part of $\ket{\xi}\bra{\xi}$, which replaces the 
$\max$ nodes at each level $i$ by the weighted sum using $a_i$.
That is
\beq
\Tr[Z_{n+1} \diag(\ket{\xi}\bra{\xi})] = 
\Tr_{a_1}\Tr_{a_2}\Tr_{a_3}\cdots\Tr_{a_n} Z_{n+1}.
\eeq
\noindent
This will turn out to be the wrong choice for $W$ but it gets us closer to the
following lemma:

\begin{lemma}
Let $\ket{\Psi}$ be a state, not necessarily normalized, 
and let $D$ be the diagonal part of $\ket{\Psi}\bra{\Psi}$. 
Let $E=E^2$ be a diagonal projector.

The minimum of $\Tr(Z D)$ over diagonal real matrices $Z$,
subject to the constraint $Z \geq 2 E \ket{\Psi}\bra{\Psi} E$, is given by
$2 |\bra{\Psi}E\ket{\Psi}|^2$ and is attained by
\be
Z =  2 \bra{\Psi}E\ket{\Psi} E.
\ee
\end{lemma}

\begin{proof}
Because $Z$ is diagonal we can write $\Tr(Z D)$ as $\bra{\Psi}Z\ket{\Psi}$.
Clearly if $Z\geq 2 E \ket{\Psi}\bra{\Psi} E$ then
$\Tr(Z D) = \bra{\Psi}Z\ket{\Psi} \geq 2 |\bra{\Psi}E\ket{\Psi}|^2$. 
It is also attainable using $Z =  2 \bra{\Psi}E\ket{\Psi} E$ which satisfies 
the inequality constraint because by Cauchy-Schwarz 
$\bra{\Phi}E\ket{\Psi}\bra{\Psi}E\ket{\Phi}\leq
 \bra{\Psi}E\ket{\Psi}\bra{\Phi}E\ket{\Phi}$ for any $\ket{\Phi}$.
\end{proof}

Because $\ket{\xi_B}\bra{\xi_B} = 2 E_1 \ket{\xi}\bra{\xi} E_1$, we are
almost in the situation covered by the above lemma.
Unfortunately, we are only maximizing over the space consistent
with the entries to every $\max$ node being equal (while keeping the zero
entries in $Z_{n+1}$ equal to zero), so in general $Z_{n+1}$ proportional to 
$E_1$ is not a valid solution. However, by rescaling
the variables, we can get to the situation where this subspace contains $E_1$, 
and therefore the above lemma is useful.

More specifically, let $S$ be a diagonal positive matrix. Define 
$Z_{n+1}' = \sqrt{S} Z_{n+1} \sqrt{S}$. We now can minimize 
$\beta = \Tr(Z_{n+1}' \sqrt{S^{-1}} W \sqrt{S^{-1}})$ subject to the 
constraint $Z_{n+1}' \geq \sqrt{S} \ket{\xi_B}\bra{\xi_B} \sqrt{S}$. 
We would like to choose $S$ so that $Z_{n+1}'=E_1$ is a valid solution,
that is, when $Z_{n+1}=\sqrt{S^{-1}}E_1\sqrt{S^{-1}} = S^{-1} E_1$
is put into the Sum-Max tree, the pair of values entering each $\max$ node 
are equal. We also need to define $W$ as the diagonal part of
$S  \ket{\xi}\bra{\xi} S$. For this to be valid, we must show that
we can compute $\beta$ as a function of $Z_{n+1}$ by the expression 
$\Tr(Z_{n+1} W)$ for every $Z_{n+1}$ consistent with the original requirements.
If these two conditions are satisfied, though, 
then the lemma tells us that
\be
\beta = 2 \lp| \bra{\xi} S E_1 \ket{\xi} \rp|^2,
\ee
\noindent
where we used the fact that both $S$ and $E_1$ are diagonal in the 
computational basis.

We begin by analyzing as an example the case of $n=3$ depicted in 
Fig.~\ref{fig:beta}. Define $e_i = \bra{i}E_1\ket{i}$ which takes the values
zero or one. Similarly, let $s_i = \bra{i}S\ket{i}$. We construct $S$ so that
\be
s_0=s_1 &=& \sigma_{0} \sigma_{0L}, \\
s_2=s_3 &=& \sigma_{0} \sigma_{0R}, \\
s_4=s_5 &=& \sigma_{1} \sigma_{1L}, \\
s_6=s_7 &=& \sigma_{1} \sigma_{1R}.
\ee

\noindent
The factors $\sigma_{0}$, $\sigma_{0L}$, and $\sigma_{0R}$ should be
thought of as being associated with the left $\max$ node. The first one
is a normalization factor, and the other two will be used to balance
the values of the left and right descendants. Similarly, the other three 
variables are associated with the right $\max$ node.

To satisfy the first constraint, we set $Z_{n+1}= S^{-1}E_1$, 
or equivalently, $x_i= s_i^{-1} e_i$. Note that the $e_i$ factor will force 
the appropriate $x_i$ variables to be zero. We focus on the left $\max$ node. 
The value entering through the left descendant is
\be
a_3 x_0 + (1-a_3) x_1 = 
\frac{a_3 e_0 + (1-a_3) e_1}{\sigma_{0} \sigma_{0L}},
\ee
\noindent
whereas entering on the right side is
\be
a_3 x_2 + (1-a_3) x_3= 
\frac{a_3 e_2 + (1-a_3) e_3}{\sigma_{0} \sigma_{0R}}.
\ee
\noindent
For the two values to be equal, we can choose 
$\sigma_{0L}=a_3 e_0 + (1-a_3) e_1$ and $\sigma_{0R} = a_3 e_2 + (1-a_3) e_3$.
Similarly, the constraint at the other $\max$ node can me met by
choosing $\sigma_{1L}= a_3 e_4 + (1-a_3) e_5$ and 
$\sigma_{1R}=a_3 e_6 + (1-a_3) e_7$.

Now we need to check the constraint on $W=\diag(S \ket{\xi}\bra{\xi} S)$.
Now $\Tr(Z_{n+1} W)$ can be described as the Max-Sum tree in 
Fig.~\ref{fig:beta}, with the max nodes replaced by sums. 
Focusing again on the left $\max$ node, 
it adds $a_2 \sigma_0^2 \sigma_{0L}^2$ of its left descendant plus 
$(1-a_2) \sigma_0^2 \sigma_{0R}^2$ of the right descendant.
We need these quantities to sum to one, and therefore
$\sigma_{0} = [a_2 \sigma_{0L}^2 + (1-a_2) \sigma_{0R}^2]^{-1/2}$. 
Similarly, we choose 
$\sigma_{1} = [a_2 \sigma_{1L}^2 + (1-a_2) \sigma_{1R}^2]^{-1/2}$ 
to normalize the sum replacing the right $\max$ node.

Now we can finally evaluate $\beta = 2 \lp|\bra{\xi} S E_1 \ket{\xi}\rp|^2$.
This can also be represented by a tree similar to the one in 
Fig.~\ref{fig:beta}, with the $\max$ nodes replaced by different sums 
as follows: the left $\max$ 
evaluates to $a_2 \sigma_0 \sigma_{0L}$ times the input from the left plus
$(1-a_2) \sigma_0 \sigma_{0R}$ times the right input. But the left and right 
inputs are respectively equal to $a_3 e_0 + (1-a_3) e_1 = \sigma_{0L}$ and 
$a_3 e_2 + (1-a_3) e_3 = \sigma_{0R}$, so the node 
evaluates to
\beq
a_2 \sigma_0 \sigma_{0L}^2 + (1-a_2) \sigma_0 \sigma_{0R}^2 = 
\sqrt{a_2 \sigma_{0L}^2 + (1-a_2) \sigma_{0R}^2},
\eeq
\noindent
which can be though of as the weighted root mean square of the values
of the two descendant nodes. The same thing happens at the right $\max$ node.
The complete expression then becomes
\begin{widetext}
\be
\beta &=&
2 \bigg\{ a_1 \sqrt{a_2 \lp[ a_3 e_0 + (1-a_3) e_1 \rp ]^2  
                + (1-a_2) \lp[ a_3 e_2 + (1-a_3) e_3 \rp ]^2 } \nonumber\\
&&
\ \ \ \ \ \  +  (1-a_1) \sqrt{a_2  \lp[ a_3 e_4 + (1-a_3) e_5 \rp ]^2
                + (1-a_2) \lp[ a_3 e_6 + (1-a_3) e_7 \rp ]^2 } \bigg\}^2
\nonumber\\
&=& 2 \bigg\{ a_1 \sqrt{a_2 a_3^2 + (1-a_2)} + (1-a_1) a_3 \bigg\}^2.
\ee
\end{widetext}

The above has the shorthand notation given by
\be
\beta = 2 \lp(\Tr_{a_1} \RMS_{a_2} \Tr_{a_3} E_1 \rp)^2
\ee 

\noindent
where we define $\RMS_{a_i}$ only on diagonal matrices, as
a weighted root mean square of eigenvalues whose basis vectors differ only
on qubit $i$. This is in the same spirit as $\Tr_{a_i}$ which does a regular 
weighted average.

For completeness, we also give the expression for $\alpha$ when $n=3$,
which can be obtained from the formulas derived below:
\be
\alpha = 2 (1-a_3) \lp( a_1 a_2^2 + (1-a_1) \rp).
\ee

The general case is almost identical. Consider the original Sum-Max tree
for a given odd $n$. Let $\mathcal{M}$ be the set of
$\max$ nodes of the original tree, that is, the set of binary nodes
with odd depth (where we define the depth of the root node as zero). 
For each $\mu\in\mathcal{M}$ we
introduce three variables: $\sigma_\mu$, $\sigma_{\mu L}$ and $\sigma_{\mu R}$,
which are to be associated with the corresponding $\max$ node. We define
the components of $S$ in terms of these variables as follows: the value 
of $s_j$, which is to be associated with leaf $j$, is given as the product
of $\sigma_\mu\sigma_{\mu L}$ for every node $\mu$ of which $j$ is a 
left descendant, times the product of $\sigma_\mu\sigma_{\mu R}$ 
for every node $\mu$ of which $j$ is a right descendant.

The conditions on $W$ are always satisfied by choosing
$\sigma_{\mu} = [a_\mu \sigma_{\mu L}^2 + (1-a_\mu) \sigma_{\mu R}^2]^{-1/2}$
for every $\mu\in\mathcal{M}$, where in a slight abuse of notation 
$a_\mu$ is the parameter associated with $\mu$
(i.e., $a_\mu = a_{d(\mu)+1}$, where $d(\mu)$ is the depth of node $\mu$).
The next condition that needs to be checked is that, when the diagonal entries
of $S^{-1}E_1$ are placed on the leaves of the original tree, the left and 
right descendants of each $\max$ node must be equal. We shall choose 
the values of $\sigma_{\mu L}$ and $\sigma_{\mu R}$ in order to guarantee 
this, in a process that begins at the lower nodes and proceeds upwards. At 
the lowest level they are chosen so that 
$\sigma_{\mu L}=a_n e_{\mu L L} + (1-a_n) e_{\mu L R}$,
where $e_{\mu L L}$ and $e_{\mu L R}$ are respectively the left and right leaf
values under the left child of node $\mu$. Similarly, we also set 
$\sigma_{\mu R}= a_n e_{\mu R L} + (1-a_n) e_{\mu R R}$, in terms of the
leaves under the right leg of node $\mu$. 
Having made such a choice, the value of node $\mu$ in the Sum-Max tree 
equals $\sigma_\mu^{-1}$ (up to multiplication by $\sigma$ factors from 
higher nodes). For every other $\mu\in\mathcal{M}$
that is not associated with the lowest level $\max$ nodes, we set
$\sigma_{\mu L}=a_{d(\mu)+2} \sigma_{\mu'}^{-1} + (1-a_{d(\mu)+2}) 
\sigma_{\mu''}^{-1}$, where $\mu'$ and $\mu''$ are respectively the 
left and right $\max$ nodes located under the left leg of $\max$ node $\mu$. 
With an equivalent choice
for $\sigma_{\mu R}$, the value entering either leg of $\max$ node $\mu$
will be $\sigma_{\mu}^{-1}$ (up to $\sigma$ factors from higher nodes)
and the second condition will be satisfied.

Finally we need to evaluate $\bra{\xi}S E_1\ket{\xi}$. Once again this is
to be done as a tree, with binary nodes corresponding to sums. The factors
of $\sigma_\mu$, $\sigma_{\mu L}$, and $\sigma_{\mu R}$ can be moved up the 
tree so that the node $\mu$ becomes the weighted sum of 
$a_\mu \sigma_\mu \sigma_{\mu L}$ times the left descendant plus 
$(1-a_\mu) \sigma_\mu \sigma_{\mu R}$ times the right descendant. 
All that remains on the leaves are the zero or
one values of $E_1$. The tree can be evaluated recursively from the bottom
up, in which case it is easy to see that the value of a non-root binary node 
outside of $\mathcal{M}$ (i.e., one of the original sum nodes), is equal to
either $\sigma_{\mu L}$ or $\sigma_{\mu R}$, where $\mu$ denotes its 
parent node, depending on whether it is a left or right
descendant respectively. On the other hand, for $\mu\in\mathcal{M}$, node
$\mu$ has value 
$a_\mu \sigma_\mu \sigma_{\mu L}^2 + 
(1-a_\mu) \sigma_\mu \sigma_{\mu R}^2 = \sigma_\mu^{-1}$.
The root node (which originally was a sum node) has value
$a_1 \sigma_{\mu'}^{-1} + (1-a_1) \sigma_{\mu''}^{-1}$, where $\mu'$ and 
$\mu''$ are respectively its left and right descendants. 
The square of this quantity multiplied by two is the
the value of our upper bound, which can be expanded using the definitions
for the $\sigma$ variables to obtain:
\be
\beta = 2 \lp(\Tr_{a_1} \RMS_{a_2} \Tr_{a_3} \RMS_{a_4} 
\cdots \Tr_{a_n} E_1 \rp)^2,
\ee 
\noindent
valid only for $n$ odd, to which the above discussion was restricted.
Though the case of even $n$ can also be found similarly, it can be obtained
from the above formula by the following observation: the protocol with
$n$ steps and constants $a_1,\dots,a_n$ is equivalent to the protocol
with $n+1$ steps and constants $a_1',\dots,a_{n+1}'$ with $a_{n+1}'=0$
and $a_i'=a_i$ for all $1\leq i\leq n$. Furthermore, if $E_1$ is the projector
associated with the $n$ step protocol, and $E_1'$ is the projectors associated
with the $n+1$ step protocol, the two matrices are related by
$\Tr_{a_{n+1}=0} E_1' = \RMS_{a_{n+1=0}} E_1' = E_1$. Therefore, for even
$n$ we have
\be
\beta = 2 \lp(\Tr_{a_1} \RMS_{a_2} \Tr_{a_3} \RMS_{a_4} 
\cdots \RMS_{a_n} E_1 \rp)^2.
\ee 

All that remains is to analyze the case where Bob is honest and Alice 
is cheating.
Though this could be analyzed using the methods presented in this section,
we can exploit further symmetries of the protocols to obtain the result.
In particular, the protocol with $n$ steps and constants $a_1,\dots,a_n$ 
is equivalent to the protocol with $n+1$ steps and constants 
$a_1',\dots,a_{n+1}'$ with $a_{1}'=1$, $a_{i+1}'=a_i$ for all $1\leq i\leq n$,
and Alice's and Bob's roles switched. Furthermore, if $E_0$ is
a projector associated with the $n$ message protocol, and $E_1'$ the
projector  associated with the $n+1$ step protocol, the two matrices 
are related by $\Tr_{a_{1}=1} E_1' = E_0$. We also need to use the fact
that $\Tr_{a_{1}=1}$ commutes through all the $\RMS$ operators because
it is a projector onto a subspace rather than a trace. Combining all the
results, we have proven the following theorem:

\begin{theorem}
In the protocol described in Section~\ref{sec:proto}, Alice's and Bob's
ability to win by cheating are upper bounded by
\be
P_A^* &\leq& \alpha = 2 \lp(\RMS_{a_1} \Tr_{a_2} \RMS_{a_3} \Tr_{a_4} 
\cdots \Tr_{a_n} E_0 \rp)^2, \nonumber\\\nonumber
P_B^* &\leq& \beta = 2 \lp(\Tr_{a_1} \RMS_{a_2} \Tr_{a_3} \RMS_{a_4} 
\cdots \RMS_{a_n} E_1 \rp)^2,
\ee
\noindent
when $n$ is even, and by
\be
P_A^* &\leq& \alpha = 2 \lp(\RMS_{a_1} \Tr_{a_2} \RMS_{a_3} \Tr_{a_4} 
\cdots \RMS_{a_n} E_0 \rp)^2 \nonumber\\\nonumber
P_B^* &\leq& \beta = 2 \lp(\Tr_{a_1} \RMS_{a_2} \Tr_{a_3} \RMS_{a_4} 
\cdots \Tr_{a_n} E_1 \rp)^2,
\ee
\noindent
when $n$ is odd.
\end{theorem}

Note that all the above formulas are valid only when the parameters 
$a_1,\dots,a_n$ are chosen so that the honest probability of winning is
$1/2$. This is the source of the factor of $2$ appearing in front
of the expressions, which could be replaced by one over the honest probability
of winning for more general scenarios.

In fact, the above formulas are even more general, as they apply to any
choice of $\{E_0,E_1\}$ as long as they are diagonal projectors. In such
a case, the symmetries that were used above are no longer valid, but direct 
computations should lead to the same formulas.

\section{\label{sec:num}Choosing $a_1,\dots,a_n$}

Recall that any choice of parameters $a_1,\dots,a_n$ subject to the
constraint $\bra{\psi} E_i \otimes E_i \ket{\psi} = \frac{1}{2}$ describes
a valid quantum weak coin-flipping protocol. Furthermore, we have an
analytic upper bound on the bias given by
\be
\epsilon \leq \max\lp(\alpha,\beta\rp) - \frac{1}{2},
\ee

\noindent
expressed as a function of these parameters. Now we need to chose values
for the parameters, which ideally should be selected to produce a bias as
small as possible.

Because the expressions for $\alpha$ and $\beta$ are complicated,
we shall employ numerical minimization to find optimal values for the 
parameters for certain small values of $n$. The values obtained will
prove the existence of protocols with the quoted biases.

Fortunately, the quality of the minimization does not need to be verified. 
For example, it would be perfectly acceptable if rather than finding the true
minimum, we only found a local minimum, or even if the outputted parameters
did not constitute a minimum at all. All that is needed is for the parameters
to satisfy the constraint, and produce the quoted bias when substituted into 
the expressions for $\alpha$ and $\beta$.

Note there is an issue with the constraint because it can only be satisfied
to the accuracy with which the parameters are specified. That is, when
the parameters are described to finite accuracy, the coin will not be
exactly fair when both players are honest. Of course, this is to be expected
for any practical implementation of the protocol. However, we also claim
that from a theoretical perspective, there are parameters close to the ones 
quoted that satisfy the constraint exactly, and produce protocols
with a bias equal to the quoted numbers to the given accuracy.

In addition to the constraint 
$\bra{\psi} E_i \otimes E_i \ket{\psi} = \frac{1}{2}$, the minimizations
were carried out with the constraint $\alpha=\beta$.
For $n=3$, we find $\alpha=\beta \simeq 0.69905$ at
$a_1 = 0.74094$, $a_2 = 0.479696$ and $a_3 = 0.186312$. Though strictly
speaking we should write that there exists a protocol with $n=3$ and
$\epsilon \leq 0.1991$, for simplicity we will write $\epsilon = 0.199$
which is understood to be correct only up to the given accuracy.

Though so far we have only derived an upper bound on the bias,
it is fair to claim that there exist protocols with a bias equal to
the upper bound because protocols can always be weakened. For example, 
in the first round with some probability Alice decides to let Bob determine
the outcome of the coin, otherwise, with some probability Bob decides to
let Alice determine the coin outcome and, if none of these events take place,
then the protocol is started normally.

Continuing with the analysis of small $n$, we find: for $n=4$
we get $\epsilon\simeq 0.1957$, for $n=6$ we get 
$\epsilon\simeq 0.1937$, for $n=8$ we get $\epsilon\simeq 0.1931$, and
for $n=10$ we get $\epsilon \simeq 0.1927$. For completeness, we list
the values used for $n=8$: $a_1= 0.680706$, $a_2=0.43281$, $a_3=0.323787$,
$a_4=0.264123$, $a_5=0.224377$, $a_6=0.197997$, $a_7=0.177191$ and
$a_8=0.0834815$.

To analyze larger values of $n$ one needs to change the exponential formulas
for $\alpha$ and $\beta$ into expressions that can be computed in a time
linear in $n$. This can be done because of the special structure of
our choice of $\{E_0,E_1\}$. Because the construction is not central to the
claims of this paper, we shall only give a brief example below:

The expression
$\Tr_{a_1} \RMS_{a_2} \cdots \Tr_{a_n} E_1$
can be computed using a tree, with binary nodes alternating between
weighted sums and weighted root mean square. The leaves at the lowest
level take only the two values zero and one, which should be though of as
a high value and a low value. Entering into the lowest sum nodes
are only two possibilities: both entries are the high value, or
the left descendant is high and the right one low. 
Therefore, if we assign values to the 
sum nodes they will only take on two values, a high value (equal to the 
previous high value) and a low value (equal to the weighted sum 
$a_n \text{High} + (1-a_n) \text{Low}$). Entering into the next
level RMS node there are also only two possibilities: two low values,
or a high (from the right) and a low (from the left). 
This structure repeats all the way up to the root node,
which is computed as the weighted sum of the high and low of the previous node.

In summary, the value of 
$\Tr_{a_1} \RMS_{a_2} \cdots \Tr_{a_n} E_1$ can be computed in linear
time as follows: Start with $(H_n,L_n)=(1,0)$ and update using the two rules
(1) if $n-i$ is even then $H_{i-1}= H_i$ and $L_{i-1}=a_i H_i + (1-a_i) L_i$,
(2) if $n-i$ is odd then $L_{i-1}= L_i$ and 
$H_{i-1}=\sqrt{a_i L_i^2 + (1-a_i) H_i^2}$. The value of the root node is
$L_0 = a_1 H_1 + (1-a_1) L_1$. The value of $\beta$ is then two times the 
value of the root node squared. The value of the constraint can also
be computed in a similar way by replacing the weighted root mean square
average with a weighted linear average.

\begin{figure}[tb]
\includegraphics[scale=.75]{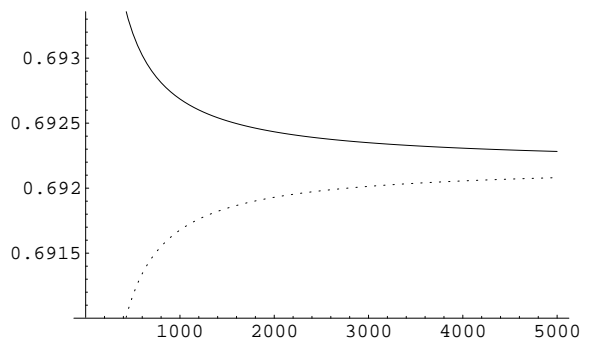}
\caption{Plot of the bounds on cheating as a function of $n$, for
parameters $a_k=1/k$. The solid line corresponds to $\beta$ and 
dotted line to $\alpha$. These bounds are only valid for even $n$.}
\label{fig:plot}
\end{figure}

Using these linear time formulas, it is possible to compute $\alpha$ and
$\beta$ for large values of $n$ given a specific functional form for $a_k$
as a function of $k\leq n$. A theoretically pleasant, though non-optimal 
choice, is $a_k = 1/k$ for $n$ even. 
Recall that $\{E_0,E_1\}$ can be described by the process
whereby the qubits are examined starting from qubit $n$ to qubit $1$,
and the first zero that is found determines the winner. With $a_k = 1/k$,
the probability that qubit $k$ needs to be examined is $k/n$, and therefore
each qubit determines the outcome with a probability of $1/n$. The problem
with this choice is that Bob's probability of winning given that qubit
$k$ needs to be examined keeps oscillating between $1/2$ and numbers greater
than $1/2$. That is why for $a_k = 1/k$, Bob has the ability to cheat
a lot, whereas Alice is more restricted. With some work, the problem
could be fixed by adjusting the values of $a_k$ for $k\sim 1$ and $k\sim n$.

The values of $\alpha$ and $\beta$ as a function of even $n$ for $a_k = 1/k$
have been plotted in Fig.~\ref{fig:plot}. The upper
solid line corresponds to $\beta$ and the lower dotted line to $\alpha$. 
For odd $n$ we would have to use $a_k = 1/(k+1)$ to satisfy the constraint,
and this would switch the values of $\alpha$ and $\beta$.

For large $n$, the graphs converge towards $0.6922$, or a bias of 
$\epsilon=0.1922$. The same behavior occurs with many other reasonable
choices for $a_k$ as a function of $k$. We believe that all choices
will converge in the limit of $n\rightarrow\infty$ to a bias of $0.1922$
(or higher for bad choices), but we shall not prove this in the present paper.

\section{Summary}

We have shown the existence of quantum weak coin-flipping protocols with
biases as low as $0.193$ and converging to a number near $0.192$ as 
$n\rightarrow \infty$. Unfortunately, this appears to be the smallest
bias that can be achieved by the protocol described in this paper for any
number of rounds.

Many possibilities remain open. The first is that quantum weak coin-flipping
with arbitrarily small bias is impossible, and that the optimal bias is either
$0.192$ or some number below it. This would be unfortunate, but in the 
opinion of the author not all that improbable.

Another possibility is that there exists a different family of protocols
that produces arbitrarily small bias. In fact, a third possibility is that
the protocol presented in this paper has an arbitrarily small bias.
This could happen either because some better choice of parameters 
$\{a_k\}$ does produce arbitrarily small bias, or because the upper bounds
$\alpha$ and $\beta$ are not tight and converge to a different value
than the true bias. The last possibility could be eliminated by constructing
cheating strategies that achieve biases equal to the upper bound.

Further research will be needed to distinguish these possibilities, and
to finally settle the question of whether quantum weak coin-flipping 
with arbitrarily small bias is possible.

\begin{acknowledgments}
 
The author would like to thank Ben Toner, Alexei Kitaev and John Preskill 
for their help. This work was supported in part by
the National Science Foundation under grant number EIA-0086038 and
by the Department of Energy under grant number DE-FG03-92-ER40701.
 
\end{acknowledgments}


\end{document}